\documentclass[epsfig,12pt]{article}
\usepackage{epsfig}
\usepackage{graphicx}

\newcommand{\beq}{\begin{equation}}   
\newcommand{\eeq}{\end{equation}}

\newcommand{\gsim}{\lower.7ex\hbox{$
\;\stackrel{\textstyle>}{\sim}\;$}}
\newcommand{\lsim}{\lower.7ex\hbox{$
\;\stackrel{\textstyle<}{\sim}\;$}}

\begin{document}

\begin{titlepage}

\begin{flushright}
FTPI-MINN-11/33, UMN-TH-3025/11\\
\end{flushright}

\vspace{0.7cm}

\begin{center}
{  \large \bf  Degeneracies in  Supersymmetric  Gluodynamics \\[2mm] and its Orientifold Daughters at large \boldmath{$N$}}
\end{center}
\vspace{0.6cm}

\begin{center}

 {\large
    M. Shifman}
\end {center}

\begin{center}

{\it  William I. Fine Theoretical Physics Institute, University of Minnesota,
Minneapolis, MN 55455, USA}\end{center}

\vspace{0.7cm}

\begin{center}
{\large\bf Abstract}
\end{center}

\hspace{0.3cm}
	I use the Nicolai map and ensuing (super)locality of appropriate correlation functions
	to prove the existence of an infinite number of
degeneracies in the mass spectra and decay coupling constants in supersymmetric  gluodynamics   
and its daughter orientifold theory at large $N$.
	
\vspace{2cm}

\end{titlepage}

\vspace{5mm}


In this paper an infinite number of parity 
degeneracies of the mass spectra and decay coupling constants will be shown to
exist in supersymmetric  gluodynamics   and its daughter orientifold theory at large $N$.
For instance, the masses of the glueballs with $J^P = k^\pm$ ($k=0,1,2,...$) are degenerate.

Consider the simplest ${\mathcal N}=1$ supersymmetric Yang--Mills theory (referred to as supersymmetric gluodynamics). 
For definiteness we
assume the gauge group to be SU$(N)$. The Lagrangian includes the gluon and gluino fields,
\begin{eqnarray}
{\mathcal L} 
&=&  \left\{ \frac{1}{4g^2}\, 
\int d^2 \theta\, \,W^{a\,\alpha} \, W_\alpha^a \  +{\rm H. c.}\right\} 
\nonumber\\[3mm]
&=& -\frac{1}{4g^2}F_{\mu\nu}^a\, F_{\mu\nu}^a
+\frac{i}{g^2}\lambda^{a\,\alpha}{\mathcal D}_{\alpha\dot\beta}
\bar \lambda^{a\,\dot\beta}\,,
\label{onesg}
\end{eqnarray}
(this is in Minkowski space, for a review see \cite{shi}). This theory is supposed to be confining,
i.e. its physical spectrum consists of color-singlet mesons and baryons, e.g. glueballs.

The basic object of our analysis is a  local {\em single-trace} operator
\beq
{\mathcal O}_{\alpha_1\alpha_2 ...} = {\rm Tr} \left ( {\mathcal F}\, {\mathcal F}\, ... \right)_{\alpha_1\alpha_2 ...} - {\rm Lorentz \,\, traces}
\label{ohs}
\eeq
where ${\mathcal F}$ is the self-dual gluon field strength tensor,
\beq
{\mathcal  F_{\mu\nu}} \equiv F_{\mu\nu} +i \tilde{F}_{\mu\nu}\,,
\eeq
and the operator ${\mathcal O}_{\alpha_1\alpha_2 ...}$ polynomially depends on a number of the gluon self-dual operators
${\mathcal F}_{\alpha\beta}$.  
None of the Lorentz indices are assumed to be contracted. Moreover, we will assume
that the operator ${\mathcal O}_{\alpha_1\alpha_2 ...}$ has the maximal possible Lorentz spin
compatible with its composition. Since ${\mathcal  F_{\mu\nu}}$ belongs to the
$(1,0)$ representation of the Lorentz group, the maximal spin produced by the operator ${\mathcal O}_{\alpha_1\alpha_2 ...} $
is $k$, where $k$ is the number of ${\mathcal F}$ factors in $ {\mathcal O}_{\alpha_1\alpha_2 ...\,}$.

\vspace{1mm}

Our consideration will consist of several steps.
 The first step is based on the Nicolai map \cite{Nicolai} in supersymmetric gluodynamics
 \cite{vene}. After performing the Nicolai mapping
the theory becomes free, i.e. the partition function can be written as the following path integral:\footnote{The path integral is
written in Euclidean space.}
\beq
\int {\mathcal D} {\mathcal  F_{\mu\nu}} \exp \left[\int  \, d^4 x\left( -\frac{1}{2g^2}\, {\rm Tr} \, { \mathcal F_{\mu\nu}}^2\right)\right]\,.
\label{4}
\eeq
In deriving Eq. (\ref{4}) one performs functional integration over the gluino fields and then, from the 
functional integration over ${\mathcal D} A$ (in the light-cone gauge) one 
 proceeds to the functional integration over ${\mathcal D} {\mathcal  F_{\mu\nu}}$. Three local variables 
 in $A_\mu (x)$ are traded for three variables residing in ${\mathcal  F_{\mu\nu}} (x)$. The functional determinant obtained from the integration over the gluino fields exactly cancels the Jacobian $\delta A(x)/\delta {\mathcal  F}(y)$. The Nicolai mapping
 and its applications were rarely discussed (if at all) in the  last two decades. A revival of interest is due to
 the recent  paper \cite{boch}.

\begin{figure}[h!t]
\begin{center}
\epsfxsize=7.0cm
 \epsfbox{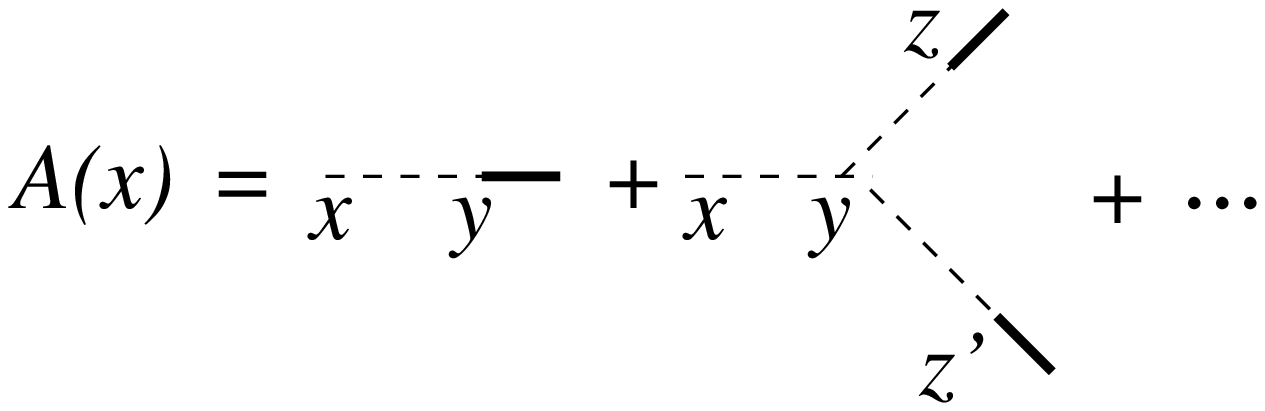}
\caption{\small A nonlocal expression for $A$ in terms of ${\mathcal  F}$. The dashed line stands for 
the nonlocal operator $\partial^{-1}$. The thick solid line denotes  ${\mathcal  F}$. }
\label{domain}
\end{center}
\end{figure}

In terms of $ {\mathcal  F}$  the gluon potential $A_\mu$ can be written as nonlocal operator 
through the inversion of the Nicolai map.
For instance, in a symbolic form
\begin{eqnarray}
A(x) &=& \int d^4y \left(x-y\right)^{-3} \left\{\rule{0mm}{5mm}{\mathcal  F}(y) \right.
\nonumber\\[3mm]
&-&
\left.
\int d^4z \,d^4 z^\prime \left(y-z\right)^{-3} \left(y-z^\prime\right)^{-3}\left[ {\mathcal  F}(z)\,,\,  {\mathcal  F}(z^\prime)
\right]+ ...
\right\}\,,
\end{eqnarray}
see Fig. \ref{domain}

In the momentum space, say, in the linear approximation $$A_\alpha = \frac{k_\beta}{k^2} {\mathcal  F_{\mu\nu}} C^{\mu\nu\beta\alpha}\,,$$ with purely numerical coefficients $C$.

The inversion of the Nicolai mapping is problematic near self-dual points where $ {\mathcal  F}$ vanishes \cite{boch}.
However, in the 't Hooft limit
self-dual field configurations are expected to be suppressed \cite{mig} in the functional 
integrals. In particular, the $\theta$ dependence is always
suppressed by $1/N$ and is not seen in the limit $N\to \infty$.
Moreover, self-dual fields are certainly
unimportant for excited states in the spectrum 
(for which predictions of the type (\ref{14}) below apply equally well).
The irrelevance of self-dual fields at $N=\infty$ in super-Yang--Mills is akin to the same observation 
done long ago by Witten in a two-dimensional example \cite{Witten}. 
This motivates us to
limit ourselves to the large-$N$ limit. Another reason for 
sticking to this limit will be indicated below. 

Summarizing,  irrelevance of the self-dual fields in the mass spectrum at $N=\infty$ must be viewed
as a well-motivated physical assumption which is hard to avoid (this is also in agreement with 
 some results in \cite{boch}). If so, all predictions for the masses and coupling constants
 following from the Nicolai map are solid.

Now, let us consider the two-point function
\begin{eqnarray}
\Pi_{\alpha_1\alpha_2   ...\alpha_1^\prime\alpha_2^\prime ...}  (q) 
&=&
 i \int d^4 \exp (iqx) \, \langle {\mathcal O}_{\alpha_1\alpha_2 ...} (x) \, 
{\mathcal O}_{\alpha_1^\prime\alpha_2^\prime ...}  (0)\rangle 
\nonumber\\[3mm]
&=&
 i \int d^4 \exp (iqx) \,\, \int {\mathcal D} {\mathcal  F} \exp \left[\int  \, d^4 x\left( -\frac{1}{2g^2}\, {\rm Tr} \, { \mathcal F_{\mu\nu}}^2\right)\right]
\nonumber\\[3mm]
&\times&
 {\mathcal O}_{\alpha_1\alpha_2 ...} (x) \, 
{\mathcal O}_{\alpha_1^\prime\alpha_2^\prime ...}  (0)\,.
\label{6}
\end{eqnarray}
In analyzing the above equation we will focus on the kinematic structure with the highest spin,
namely, 
\beq
\Pi_{\alpha_1\alpha_2   ...\alpha_1^\prime\alpha_2^\prime ...}  (q) =
\left(g_{\alpha_1\alpha_1^\prime}\, g_{\alpha_2\alpha_2^\prime}... + {\rm permutations}
\right) + ...
\label{7}
\eeq
Since the functional integration in (\ref{6}) runs over ${\mathcal D} {\mathcal  F}$, treated as independent variables,
it is obvious that the correlation function (\ref{6}) will vanish unless $x=0$.
More exactly,\footnote{The pre-exponent in Eq.~(\ref{8}) is not gauge gauge invariant. One should understand this
relation  somewhat symbolically, as a building block for  two-point functions with the pre-exponent composed of expressions (\ref{ohs})
which are gauge invariant. }
\begin{eqnarray}
&&\int {\mathcal D} {\mathcal  F} \,\,  {\mathcal  F_{\mu\nu}} (x) {\mathcal  F_{\rho\sigma }} (y)
\exp \left[\int  \, d^4 x\left( -\frac{1}{2g^2}\, {\rm Tr} \, { \mathcal F_{\mu\nu}}^2\right)\right]
\nonumber\\[3mm]
&&
\propto \left(g_{\mu\rho} g_{\nu\sigma} - g_{\nu\rho} g_{\mu\sigma}
\right)\delta^4 (x-y)\,.
\label{8}
\end{eqnarray}
We will refer to this property of the specifically designed correlation functions 
as to superlocality.

The same is valid for all operators ${\mathcal O}$ which are polynomially expressed in terms of
${\mathcal F}$.
Since the action is quadratic in ${\mathcal  F}$, and the same ${\mathcal  F}$ is the functional integration variable,
in evaluating (\ref{6}) one can use the pairwise Wick contraction (after expressing the operators ${\mathcal O}_{\alpha_1\alpha_2   ...}$   in  terms of ${\mathcal  F}$'s).

If so, upon the functional integration the correlation function (\ref{6}) contracts and becomes
 a (generalized) tadpole, with no imaginary part,
\beq
{\rm Im}\,\Pi_{\alpha_1\alpha_2   ...\alpha_1^\prime\alpha_2^\prime ...}  (q) =
\left(g_{\alpha_1\alpha_1^\prime}\, g_{\alpha_2\alpha_2^\prime}... + {\rm permutations}
\right) \times {\rm zero}\,.
\label{9}
\eeq
Equation (\ref{9}) should be trivial in perturbation theory. However, let us not forget that
supersymmetric gluodynamics confines, and its physical spectrum at large $N$ consists of massive 
stable mesons  with
certain spins and parities\,\footnote{At this point the large-$N$ limit is useful but not crucial.
At finite $N$ the spectrum is no longer represented by
an infinite sum of poles. Instead, the resonances acquire finite widths, and multiparticle states contribute to the
two-point correlation function (\ref{6}).} 
(there is no massless states in the physical spectrum). Then (\ref{9})
imposes constrains on the spectra.

Let us examine these constrains first in the example of, say, cubic in ${\mathcal F}$ operators.
Then the operator ${\mathcal O}$ takes the form
\beq
 {\mathcal O} \to {\rm Tr}\,  \left(E^{k_1}-iB^{k_1}\right)  \left(E^{k_2}-iB^{k_2}\right) \left(E^{k_3}-iB^{k_3}\right) 
\eeq
where 
$E$ and $B$ are chromoelectric and chromomagnetic fields, respectively, and $k_{1,2,3} = 1,2,3$. Complete symmetrization over
$k_1, k_2, k_3$ is assumed. It can be split in two parts with definite parities,
\begin{eqnarray}
{\mathcal O}_- &=&
 {\rm Tr}\,   E^{k_1}\,    E^{k_2}\, E^{k_3} - \left( {\rm Tr}\,   B^{k_1}\,    B^{k_2}\, E^{k_3}
+ {\rm perm.}\right),
\nonumber\\[2mm]
{\mathcal O}_+ &=&
  {\rm Tr}\,   B^{k_1}\,    B^{k_2}\, B^{k_3} - \left( {\rm Tr}\,   E^{k_1}\,    E^{k_2}\, B^{k_3}
+ {\rm perm.}\right). 
\end{eqnarray}
The operator ${\mathcal O}_- $ creates from the vacuum $J^P=3^-$ meson (in its rest frame
$\left\langle 0 \left| {\rm Tr}\,   E^{k_1}\,    E^{k_2}\, E^{k_3} +... \right| J^P=3^-\right\rangle = {\rm const}\times\, \varepsilon^{k_1k_3k_3}$
where $ \varepsilon^{k_1k_3k_3}$ is polarization vector;  full symmetrization is assumed). A similar expression can be written for
$\left\langle 0 \left| {\rm Tr}\,   B^{k_1}\,    B^{k_2}\, B^{k_3} +... \right| J^P=3^+\right\rangle$.
The two-point function (\ref{6})
reduces to
\beq
\langle {\mathcal O}_- \,,\, {\mathcal O}_- \rangle - \langle {\mathcal O}_+ \,,\, {\mathcal O}_+\rangle
\eeq
where we take into account the fact that the $P$-parity is unbroken in supersymmetric gluodynamics, and, hence,
the cross correlator $\langle {\mathcal O}_- \,,\, {\mathcal O}_+ \rangle $ must vanish.
The imaginary part  of $\langle {\mathcal O}_- \,,\, {\mathcal O}_- \rangle$ is positive-definite, and so is
the imaginary part  of $\langle {\mathcal O}_+ \,,\, {\mathcal O}_+\rangle$.
Since the overall imaginary part vanishes, the first one should cancel the second.

Thus, we conclude that the masses and residues of positive and negative parity states of spin 3 must be degenerate.
This is obviously valid for arbitrary spins.

We observe parity degeneracy of the physical spectra produced by two distinct operators
obtained from each other by the substitution, $E \leftrightarrow B$. In addition to parity degeneracy discussed above,
we certainly have supersymmetric degeneracies, following from the fact
that massive supersymmetry representations contain three subsequent spin states, for instance (0,1/2,1/2,1)
in the vector superfield. This general degeneracy must be superimposed with the parity degeneracy specific to supersymmetric 
gluodynamics.

Degeneracies following from superlocality of the correlation
functions of the type (\ref{6}) are valid not only for the mass spectrum, but for the decay constants too.
For instance, consider a three-point correlator
\beq
\langle {\mathcal O} (x)  \,,\, {\mathcal O} (y)  \,,\, {\mathcal O} (0) \rangle  
\label{13}
\eeq
where for simplicity I  limit myself to the spin-zero operator 
${\mathcal O} = {\rm Tr}  \, {\mathcal F}^2$. Acting on the vacuum, the operator ${\mathcal O}$
produces either  scalar (S) or pseudoscalar (P) glueballs, whose masses are degenerate at every level. 
Following the same line of reasoning as above, it is easy to see that  the absence of physical cuts in (\ref{13})
implies that at  every level 
\beq
g_{SSS} = - 3\, g_{SPP}\,,
\label{14}
\eeq
where the coupling constants $g$ in the expression above are
the $S$-wave decay constants $S\to SS$ and $S\to PP$, respectively. One can consider off-diagonal transitions too
(i.e. from one level to another).

Because of the planar equivalence 
between supersymmetric gluodynamics and the orientifold theory \cite{Armoni}
 the same conclusion applies to Yang--Mills theory with one Dirac quark in the two-index
antisymmetric or symmetric representation. In the former case at $N=3$ we get just one-flavor QCD.
It would be interesting to study $1/N$ corrections to these predictions, perhaps, on lattices.

Finally,  I would like to  mention 
that
two particular results of the same nature had been established
 in the literature previously on different grounds. First,
 the 
fact of degeneracy of  spectra 
(imaginary parts) in the scalar and pseudoscalar channels, 
$\langle F^2(x) \,,\, F^2(0)\rangle$ and $\langle F\tilde F (x) \,,\, F \tilde F\rangle$, respectively,
 dates back to  \cite{Novikov}. Second, it was 
shown \cite{Gorsky}
that the spectral functions associated with the  
(nonchiral!) operator Tr $\left({\mathcal F}_{\mu\nu} \bar\lambda^2\right)$ are fully
degenerate in the $J^{PC} = 1^{\pm}$ channels. This statement follows from
${\mathcal N} = 1/2$ supersymmetry discovered in \cite{Seiberg}, to which
the above operator is related. It is probable that extensions of Seiberg's construction \cite{Seiberg}
can be worked out
providing an alternative derivation of the results presented in this note.

Verification of the predicted degeneracies seems to be an excellent testing ground for
lattice studies of supersymmetry at strong coupling, both in the Lagrangian and Hamiltonian formulations.

I am very grateful to Adi Armoni, Marco Bochicchio,  Sasha Migdal and Nati Seiberg for stimulating discussions.
This work is supported in part by DOE grant DE-FG02- 94ER-40823.

\end{document}